\documentstyle[prl,aps,floats,psfig]{revtex}

\begin{document}
\twocolumn[
\hsize\textwidth\columnwidth\hsize\csname@twocolumnfalse\endcsname

\draft
\title{
Ultrasonic attenuation in $d$-wave superconductors
}
\author{
Vladimir N. Kostur, Jayanta K. Bhattacharjee$^{a}$,
        and Richard A. Ferrell
}
\address{
Center for Superconductivity Research,
Department of Physics\\
University of Maryland, College Park, MD 20742-4111}
\date{E-print cond-mat/9506045; June 9, 1995}
\maketitle

\begin{abstract}
The longitudinal ultrasonic
attenuation  at low temperatures  for a
$d$-wave superconducting order parameter is examined.
The ratio
of the superconducting to normal state attenuation,
 $\alpha_{s}/\alpha_{n}$, depends strongly on the
direction of propagation of ultrasound
with  maxima  corresponding to nodes of the order parameter
on the Fermi surface.
We propose that measurements of ultrasonic attenuation in a single
crystal
of the high-$T_{c}$ superconductors  can give  additional information
regarding the pairing symmetry.

\end{abstract}

\pacs{PACS numbers: 74.20.-z, 74.40.+k, 74.50.+r, 74.70.Mq}
]

An important aspect of the high-$T_c$ superconductors
that is currently receiving a great deal of attention
is the angular dependence of the energy gap.
Some experimental evidence \cite{Vollman,Wellstood} has been brought
forward
for $d$-wave pairing instead of the more conventional
BCS $s$-wave pairing.
The purpose of this note
is to point out that
ultrasonic attenuation
may offer a very effective and useful tool
for the experimental determination of the variation of the energy gap
around the Fermi surface.
We demonstrate that, due to the layered structure
of the high-$T_c$ cuprates,  the anisotropy of the $d$-wave
order parameter
should result in a very  pronounced dependence of
the  ultrasonic attenuation on the
direction of propagation (parallel to the $CuO_2$ layers) at low
temperatures.

In order to concentrate on our main point,
we assume
that the scale of the quantum of sound energy, $\hbar \omega_q$,
is negligibly small
compared to the energy gap.
Typical experimental values of $\omega_q$ are
of the order of $10^{9}-10^{10}\ sec^{-1}\approx 0.01-0.1\ meV$,
equivalent to a temperature of 1 K or less. (For simplicity,
we ignore the immediate vicinity of a node.)
This  assumption leads to the picture of the ultrasonic attenuation
as a kind of quasi-elastic scattering of the quasiparticles of energy
$E$ by the incident sound quantum.
The initial states are counted in terms of increments of
the normal state energy,
$d\epsilon$ = $(\partial \epsilon/ \partial E)dE$.
The integration for the total transition rate in the Fermi
``golden rule'' formula needs to include an additional factor
$\partial \epsilon/ \partial E$ for the final state density.
But these two factors of $\partial \epsilon/ \partial E$
are cancelled by the squared  strength
of the coupling of the sonic field with the quasi-particle energy
\cite{Ferrell}, which is proportional to
$\partial E / \partial \epsilon$. This follows from the fact that it
is to $\epsilon$,  and only indirectly to $E$, that the lattice
deformation is coupled. Consequently, the ratio
of the longitudinal ultrasonic attenuation in the
superconducting and normal
state is given by the  formula
\begin{equation}
\frac{\alpha_s}{\alpha_n} =
-2 \int_{\Delta}^{\infty} \frac{\partial f(E)}{\partial E}
dE = 2f(\Delta) \equiv
\frac{2}{\exp{\beta \Delta}+ 1}\>,
\end{equation}
where $f(E)$ is the Fermi function and
$\Delta$ is the superconducting energy
gap. This derivation\cite{Ferrell}
 of Eq.~(1)
does not involve the BCS coherence factors. They do, however,
enter
for frequencies so high that
the energy of the sound quantum is not negligibly small,
 as can be seen
in the  standard treatment \cite{Tinkham}.
When $\Delta$ differs, as presently under consideration, for different
quasi-particles at the Fermi surface
we take the average of the right hand
side of Eq.~(1), weighted by the contribution in the normal state.
\begin{figure}
\centerline{\psfig{file=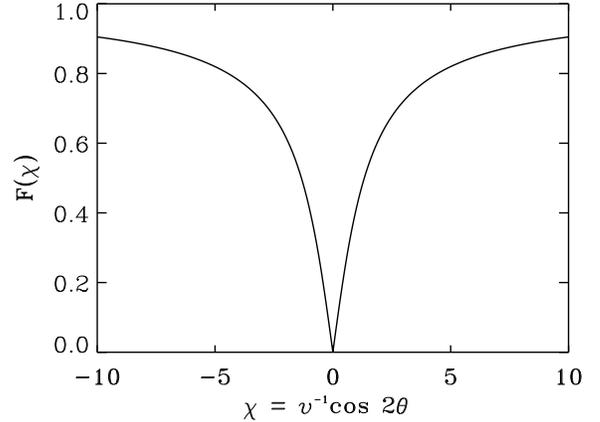,width=0.95\linewidth}}
\caption{The scaling function for the angular dependence
of the $d$-wave order parameter. This function characterizes
the universal behavior which connects  small doping and the
angle dependence
of the order parameter.}
\label{fig1}
\end{figure}
It is possible to obtain the same result using Green's functions
to calculate the density-density
correlation function,  the imaginary part of which is proportional to
ultrasonic attenuation. The  result  for
longitudinal ultrasonic attenuation in anisotropical
superconductors is given by
the following expression\cite{Pokrovskii}:
\begin{equation}
\frac{\alpha_s (\theta)}{\alpha_n (\theta)} =
\frac{ <\delta({\bf v_{\bf k} \cdot q})\  |g({\bf k, k})|^{2}
\ 2f(\Delta_{\bf k})>_{FS}}
{<\delta({\bf v_{\bf k} \cdot q})\ |g({\bf k, k})|^{2}>_{FS}}\>,
\end{equation}
where  $g({\bf k, k})$ is the matrix element of
electron-(acoustic) phonon interaction. The momentum $\bf k$
and velocity $\bf v_{\bf k}$ are
on the Fermi surface and $< ... >_{FS}$ is the
integration over the Fermi surface. The angle $\theta$
describes the direction of $\bf q$, the transfer momentum.
In the case of a two-dimensional Fermi surface, the integrations
 in Eq.~(2)
are reduced because of the $\delta$-functions and the matrix element
$g({\bf k, k})$ in  Eq.~(2)
cancells.
For concreteness,  we consider the two-dimensional tight-binding
electron spectrum
\begin{equation}
\epsilon_{\bf k} = -2t[\cos{k_x} +\cos{k_y}] -\mu \>,
\end{equation}
where $t$ is
the hopping matrix element. $\mu$ is the shift of chemical potential
as determined
by the band-filling factor $n$ (the Fermi surface is specified
by $\epsilon_{\bf k} = 0$).
The lattice constant is taken to be 1 for simplicity.
\begin{figure}
\centerline{\psfig{file=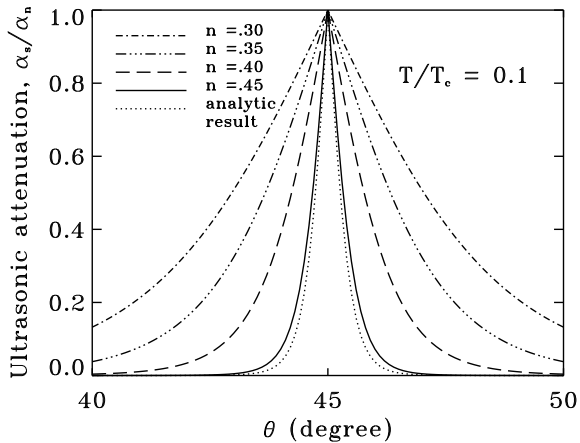,width=0.95\linewidth}}
\caption{The dependence of longitudinal
ultrasonic attenuation at $T/T_c\ =\ 0.1$ and different
band-filling factors. The dotted line shows the analytic
result at $n=\ 0.45$ (see Eq.~(5--6)).
}
\label{fig2}
\end{figure}
The $d$-wave order parameter can  be written
in the form
\begin{equation}
\Delta_{\bf k} =\frac{\Delta_{0}(T)}{2}
[\cos  k_x - \cos k_y]\>,
\end{equation}
where $\Delta_{0}(T)$
obeys the usual BCS temperature dependence.
One can see that, in any given direction, the
ultrasonic attenuation falls off exponentially
 with
temperature
except for directions
close to those of the nodes of the $d$-wave order parameter.
Exactly at  the nodes,  $\alpha_s / \alpha_n = 1$,
independent of $T$.
This yields  sharp spikes in the angular dependence
for $\alpha_s / \alpha_n$, which become more pronounced
 with decreasing temperature.
The analytic result for the case where the band-filling factor is close
to one-half
( $n \simeq 0.5 - 3\pi^{-2} \nu [1 + \ln (2/\sqrt[3]{\nu})]$,
with  $\nu = |\mu|/4t \ll 1$) is a kind of universal scaling, so
that all such small values of $\nu$ are described by the single
scaling function
\begin{equation}
F(\chi) = \frac{\chi}{1+\sqrt{1+\chi^{2}}}\>,
\end{equation}
which is plotted in Fig.\ \ref{fig1}
versus the scaling variable
$\chi =\nu^{-1} |\cos 2 \theta |$. In this case of the square lattice,
$\theta$ is the angle between $\bf q$ and the $y$-axis. Thus, Eq.~(2)
becomes
\begin{equation}
\frac{\alpha_s (\theta)}{\alpha_n (\theta)}
\simeq
2\Big[ 1 + \exp \Big(\frac{\Delta_{0}(T)}{T} F(\chi)\Big)\Big]^{-1}\>,
\end{equation}
It follows from Eqs.~(5) and
(6) that,  in the low-temperature region ($T \ll T_c$),
very sharp peaks  in $\alpha_s (\theta) / \alpha_n(\theta)$
appear at $\theta = \pm 45^{\circ}, \pm 135^{\circ}$.
The sharpness of the peaks is the result of
 the anisotropy of both the gap and the Fermi surface.
The dotted curve in Fig.\ \ref{fig2}
illustrates the application of Eqs.~(5)
and (6) at the low temperature $T/T_{c} = 0.1$
for $n = 0.45$ (or $\nu =0.065$). The solid curve is calculated
numerically for the same filling  without recourse to the small
$\nu$ approximation. The closeness of the solid and dotted
curves is an indication of the accuracy for $\nu = 0.065$,
of this approximation and of the scaling function that
follows from it. The numerical computations for other fillings
are exhibited by the other three curves in Fig.\ \ref{fig2}.
These curves clearly
demonstrate the salient feature mentioned above, namely,
that at low temperature,
and even at fillings significantly far from
half-filling,
a sharp peak can be expected in
$\alpha_s/\alpha_n$ at the sonic propagation
direction that corresponds
to the nodes in the energy gap. Figure\ \ref{fig3}
shows the development of this peak, for $n=0.4$, as the temperature
is lowered.

\begin{figure}
\centerline{\psfig{file=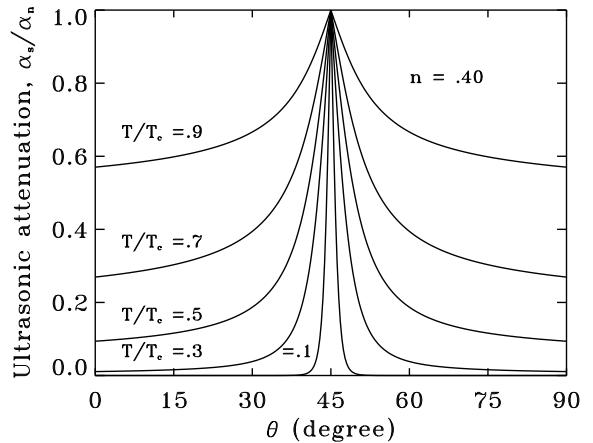,width=0.95\linewidth}}
\caption{
The angle dependence of the ratio of
longitudinal ultrasonic
attenuation in the superconducting state to the normal state
value at $n\ =\ 0.4$ and different temperatures.
 Note the exponential
decrease of ultrasonic attenuation with temperature far from nodes,
 as well
as the closeness  to one of the ratio $\alpha_s/\alpha_n$
near the nodes.
}
\label{fig3}
\end{figure}

To conclude, we note that effect of  anisotropy of the
order parameter
upon ultrasonic attenuation was  observed
in various metals
\cite{Bohm} and  in superfluid $^{3}$He \cite{Vollhardt}.
Some recent progress has been made in ultrasonic measurements
of single crystals of $UPt_3$ \cite{PhysAc}, where a
three-dimensional
$d$-wave order parameter  is assumed.
The earliest ultrasonic measurements on
high-$T_c$ crystals found a quite small velocity change
at the superconducting transition temperature,
which indicates that the influence of superconductivity
on the acoustic processes
 is not yet fully understood\cite{Allen}.
Nevertheless,
the sharp spike-like nature of the angular dependence
of the longitudinal ultrasonic attenuation
in the case of the  possible $d$-wave  order parameter
should  be
observable and should provide a clear signature
for the latter.

\newpage
The authors are grateful to G.~M.~Eliashberg for discussions.
This work is supported by NASA Grant No. NAG3-1395.


\begin{references}
\bibitem[\em a]{na}  {\em Permanent address:}
Department of Physics, Indian Institute of
Technology, Kanpur, 208016, U.~P., INDIA
\bibitem{Vollman}{D. A. Wollman, D. J. Van Harlingen, W. C. Lee,
D. M. Ginsberg, and A. J. Leggett,  Phys. Rev. Lett. {\bf 71},
2134 (1993).}
\bibitem{Wellstood}{A.~Mathai, Y.~Gim, R.~C.~Black,
A.~Amar, and  F.~C.~Wellstood, preprint.}
\bibitem{Ferrell}{
R. A. Ferrell, Physics, {\bf 3}, 157 (1967).}
\bibitem{Tinkham}{M.~Tinkham, {\em Introduction to Superconductivity},
McGraw-Hill, N. Y., 1975.}
\bibitem{Pokrovskii}{V.~L.~Pokrovskii, JETP  {\bf 40}, 898 (1961),
Soviet Phys. JETP {\bf 13}, 628 (1961).}
\bibitem{Bohm}{H. V. Bohm and N.~H.~Horwitz, in Proceedings
of the Eight International Conference on   Low-Temperature Physics,
ed. R.~O.~Davies, 1963, p.191.}
\bibitem{Vollhardt}{D.~Vollhardt and P.~W\"olfle,
{\em The Superfluid Phases of Helium 3},
Taylor \& Francis, London, 1990.}
\bibitem{PhysAc}{{\em Ultrasonics of High-$T_c$
and Other Unconventional
Superconductors}, in Physical Acoustics,
v. {\bf 20}, ed. M. Levy, Academic Presis, 1992.}
\bibitem{Allen}{P. B. Allen, Z. Fisk, A. Migliori,
in {\em Physical Properties of High Temperature
Superconductors I}, ed. D. M. Ginsberg, World Scientific 1989.}


\end{references}
\end{document}